\documentclass[a4paper,UKenglish]{mylipics}
 
\usepackage{microtype}

\usepackage{wallpaper}
\usepackage{latexsym}

\newlength{\spacelen}

\def\ms#1{\null\ifmmode\mathord{\mathcode`-="702D\it #1\mathcode`\-="2200}%
        \else$\mathord{\mathcode`-="702D\it #1\mathcode`\-="2200}$\fi}

\newcommand{\cws}[3]
        {\\[-#1pt] $$ {#3} $$ \\[-#2pt]}

\newlength{\widtharg} \newlength{\heightarg}

\newcommand{\lap}
        {\mbox{$<$}}
\newcommand{\rap}
        {\mbox{$>$}}

\newcommand{\lmp}
        {\{ \! | \,}
\newcommand{\rmp}
        {\, | \! \}}

\newcommand{\calb}
        {{\cal B}}
 
\newcommand{\caln}
        {{\cal N}}
        
\newcommand{\calp}
        {{\cal P}}

\newcommand{\calq}
        {{\cal Q}}

\newcommand{\calx}
        {{\cal X}}

\newcommand{\real}
        {\, {\rm R} \!\!\!\!\! {\rm I} \,\,\,}

\newcommand{\narrow}[1]
        {\arrow{#1}\!\!\!\!\!\!\!\!\!\!\!\!/\,\,\,\,\,\,\,\,\,\,}

\newcommand{\arrow}[1]
        {\, \auxarrow\limits^{#1} \,}

\newcommand{\auxarrow}
        {\mathop{\longrightarrow}}

\newcommand{\nil}
        {\underline 0}

\newcommand{\eqdef}
        {\buildrel \Delta \over =}

\newcommand{\infr}[2]
        {\renewcommand{\arraystretch}{1.5}
        \begin{array}{c}
        #1\\
        \hline
        #2
        \end{array}}

\newcommand{\pla}{\!\widehat{\,\lambda}}
\newcommand{\pq}{\!\widehat{q}}
\newcommand{\plap}{\!\widehat{\,\lambda'\!}}
\newcommand{\la}{\lambda}
\newcommand{\ov}{\overline}
\newcommand{\sub}{\rightarrow\!}

\theoremstyle{plain}
\newtheorem{proposition}[theorem]{Proposition}

\title{Reduction Semantics in Markovian Process Algebra
}

\author[1]{Mario Bravetti}
\affil[1]{Universit\`a di Bologna, Italy\\
INRIA, France\\
  \texttt{mario.bravetti@unibo.it}}
\authorrunning{M. Bravetti} 

\Copyright{Mario Bravetti}


\serieslogo{}
\EventShortName{}

\begin{document}

\maketitle

\begin{abstract}
Stochastic (Markovian) process algebra extend classical process algebra with probabilistic exponentially distributed time durations denoted by rates (the parameter of the exponential distribution). Defining a semantics for such an algebra, so to derive Continuous Time Markov Chains from system specifications, requires dealing with transitions labeled by rates. With respect to standard process algebra semantics this poses a problem: we have to take into account the multiplicity of several identical transitions (with the same rate). Several techniques addressing this problem have been introduced in the literature, but they can only be used for semantic definitions that do not exploit a structural congruence relation on terms while inferring transitions. On the other hand, using a structural congruence relation is a useful mechanism that is commonly adopted, for instance, in order to define semantics in reduction style for non-basic process algebras such as the $\pi-$calculus or richer ones. In this paper we show how to define semantics for Markovian process algebra when structural congruence is used while inferring transitions and, as an application example, we define the reduction semantics for a stochastic version of the $\pi-$calculus. Moreover we show such semantics to be correct with respect to the standard one (defined in labeled operational semantics style).
\end{abstract}

%
%

\section{Introduction}

The advantages of using process algebras for the performance modeling and 
evaluation of concurrent systems due to their feature of compositionality
have been widely recognized (see~\cite{Hil,BB,Pri1,Her,tcs,BD} and the 
references therein).
Particularly simple and successful has been the extension of 
standard process algebras with time delays whose duration follows
an exponential probability distribution, called Markovian
process algebras (see e.g.~\cite{Hil,BB,Pri1,Her,BHK}). 
The ``timed'' behavior of systems specified with
a Markovian process algebra can be represented by a continuous
time Markov chain (CTMC), that is a simple continuous 
time stochastic process where, in each time point, the future behavior 
of the process is completely independent of its past behavior.
Due to their simplicity CTMCs can be analyzed 
with standard mathematical techniques and software tools (see e.g.~\cite{Ste})
for deriving performance measures of systems.

The semantics of all Markovian process algebras previously introduced in the literature is defined in {\it labeled operational semantics} style, that is where transitions are labeled with actions (representing potential of communication) and transitions representing actual communications are produced by matching action transitions of parallel processes.
In this paper we present a technique that allows us, for the first time, to develop semantics for Markovian process algebra
in {\it reduction} style (see for instance the reduction semantics of $\pi$-calculus \cite{SW,MilPi}), that is where transitions are not labeled with actions and are directly produced by
readjusting the structure of terms via a {\it structural congruence relation}: such a relation is endowed
with commutative and associative laws that allow communicating processes to get syntactically adjacent to each other so
to directly produce reduction transitions. Reduction semantics is also commonly used, in that convenient for defining semantics of complex languages, see for example \cite{JO}.

Markovian process algebras extend usual process algebras with probabilistic exponentially distributed time durations denoted by rates $\lambda$ (positive real numbers), where $\lambda$ is the parameter of the exponential distribution. Defining a semantics for such algebras, so to derive CTMCs from system specifications, requires dealing with transitions labeled by rates $\lambda$. With respect to standard process algebra semantics this poses a problem: we have to take into account the multiplicity of several identical transitions, that is transitions with the same rate and source and target terms. 
The simplest way to deal with this problem (see~\cite{Hil}) is counting the number of different possible inferences of the same transition so to determine its multiplicity. 
However this can be done only for semantic definitions that do not infer transitions by means of a structural congruence relation $\equiv$ on terms, that is via the classical operational rule of closure w.r.t. $\equiv$ that makes it possible to infer a transition $P \arrow{\alpha} Q'$ from $Q \arrow{\alpha} Q'$ whenever $P \equiv Q$. 
This because, supposing that common laws like $P \equiv P | \nil$ are considered, the presence of such a rule causes the same transition to be inferred in an infinite number of ways even in the case of just one occurrence of a Markovian delay (e.g. consider $P$ just being a single Markovian delay and the congruence law above). On the other hand
using a structural congruence relation is a useful mechanism that is commonly used, for instance, when defining reduction semantics of non-basic process algebras such as the $\pi-$calculus (to compute reductions due to communication while inferring transitions, as we already mentioned).


Another similar way used in the literature to deal with Markovian transition multiplicity
is to explicitly introduce a mechanism to differently identify multiple execution of identical Markovian delays by introducing some kind of identifier that typically expresses syntactical position of the delay in the term (for example it is called a ``location'' in the case of the position with respect to the parallel operator) and whose syntax is dependent on the operators used in the language (see, for instance, \cite{GHR,Pri1}). 
Again, since Markovian transitions are, roughly, identified by indicating the path (left/right at every node) in the syntax tree of the term, this solution is not compatible with inferring transition by means of a structural congruence relation with the
commutative and associative laws (which are fundamental, e.g., to define a reduction semantics for the $\pi-$calculus). This because term structure is readjusted at need during transition inference.

Finally, techniques exploiting the representation of Markovian transitions collectively as a distribution over a target state space (a traditional approach in the context of probabilistic transition systems, see Giry functor \cite{Giry} and its generalizations, for instance \cite{SL, DEP, VR}) have been used in \cite{DLLM} and in \cite{KS} for any Markovian process algebra whose operational rules are in a Markovian extension of the GSOS rule format. 
These techniques are crucially based on collectively inferring Markovian transitions performable at a certain syntactical level
(e.g. in $P\, |\, Q$) from transitions performable at lower levels (e.g. from all Markovian transitions performable by $P$
and by $Q$). Therefore, they are, again, not compatible with using structural congruence $\equiv$ (with
commutative and associative laws) during the inference of transitions, as caused by the operational rule of closure w.r.t. $\equiv$ mentioned above, which is not in the GSOS format of \cite{KS} (the approach in \cite{KS} cannot be used in the presence of such a rule, which is fundamental in reduction semantics). Since there is not a fixed stratification of transition inference, different transitions can be inferred by readjusting the structure of terms via the commutative and associative laws of structural congruence in different ways. For instance, given a term $P | (Q | R)$ it can be that the term needs to be readjusted into $(P | Q) | R$ to infer a reduction due to a communication between $P$ and $Q$ and to $(P | R) | Q$ to infer a reduction due to a communication between $P$ and $R$. So we cannot fix a stratification of transition inference, that is first {\it collectively} inferring all transitions of $(Q | R)$ and then use them to infer all transitions of 
$P | (Q | R)$.

\subsection{The Main Idea}

In this paper we show how to define semantics for Markovian process algebras in the presence of transitions inferred by means of structural congruence, which also easily allows us to deal with a recursion operator: the possibility of expressive cycling behavior is fundamental for steady state based performance analysis.
This will make it possible to define, for the first time, the semantics of a Markovian process algebra
(we will use a simple Markovian version of $\pi-$calculus) via reduction semantics instead of labeled operational semantics, as 
done in all above mentioned approaches.

Moreover our technique is such that the presence of Markovian delays in the process algebra does not cause a significant modification of the process algebra semantics, in that most of the burden is concentrated in a couple of additional rules which are independent of the particular operators used by the algebra. More precisely, we do not have to introduce an explicit (operator dependent) mechanism to distinguish multiple occurrences of actions as in  \cite{GHR,Pri1}, or to intervene
in the semantics of every operator 
as in \cite{DLLM} or to separately define, by induction on term syntax, how rate distribution is calculated using (total) rate information attached to rules as in \cite{KS} (which also does not cope with recursion).
%
But more importantly, this opens the new possibility to introduce Markovian delays in complex languages like that of \cite{JO} without the need to completely change the semantics from reduction style to labeled style.


The idea is to introduce stochastic names (that can be seen as names for stochastic variables) and stochastic binders (binders for such names). Stochastic names are used in process algebra prefixes and numerical rates $\lambda$ are assigned by the corresponding stochastic binder. In essence (a set of) stochastic binders act as a (probabilistic/stochastic) scheduler: they use names attached to prefixes to collectively quantify them. In this way, {\it assuming that all prefixes are associated with a different stochastic name}, rate values replace stochastic names upon binding and multiplicity is correctly taken into account because prefixes with the same rates are anyhow distinguished by having different stochastic names.

But how to guarantee that prefixes are given different stochastic names? The idea is to consider a specification language where prefixes are, as usual, associated with numerical rates $\lambda$ and then to turn them (by a law of structural congruence) into stochastic names that are immediately bound with a stochastic binder assigning the rate $\lambda$. Then, with the usual rules of binder extrusion, we impose that, before binders can be evaluated, they must be all lifted to the outermost syntactical level by structural congruence. As a consequence we have the guarantee (due to the requirement of $\alpha$-conversion related to binder extrusion) that every prefix is assigned a distinguished stochastic name. Note that recursion is dealt with in the correct way as well, because, as we will see, recursions need to be unfolded in order to perform the lift of the binders. 

\subsection{Plan of the Paper}

In this paper we consider a simple extension of the $\pi-$calculus with Markovian delays, where for the sake of simplicity, Markovian delays are expressed separately from standard actions as in the Interactive Markov Chains approach of \cite{Her}.
We first define the semantics of this calculus in labeled operational semantics style (we use the early semantics of the $\pi-$calculus) according to the classical approach of \cite{GHR,Pri1}, that is using an explicit (operator dependent) mechanism to distinguish multiple occurrences of Markovian delays.
We then apply our technique based on stochastic names and binders to provide a reduction semantics for the very same
process algebra.
We finally assess the correctness of our approach by showing that a theorem similar to the Harmony Lemma in \cite{SW}
(or the one in \cite{MilPi}) holds true, that is: stochastic and standard reductions of reduction semantics are in correspondence with $\tau$ and Markovian delay transitions of the labeled operational semantics and, thus, 
the underlaying Interactive Markov Chains.

In particular, in Sect. 2 we present the Markovian $\pi-$calculus and its labeled operational semantics defined according to the ``classical'' technique. In Sect. 3 we present the reduction semantics for the Markovian $\pi-$calculus. In Sect. 4 we conclude the paper by further discussing related work  
and future extensions of the technique. Finally,
Appendix \ref{App} includes a long proof for a proposition (whose proof structure is described in the paper body).
The work in this paper is based on the technical report \cite{TechRep}.

\section{Classical Approach}\label{Classical}

We consider $\pi-$calculus with constant definitions (as in \cite{MPW}) extended with Markovian delay prefixes (following the approach in \cite{Her}). Markovian delays are thus simply represented by $(\la)$ prefixes, where $\la$ is the  
rate of a Markovian distribution.
As usual in the context of Markovian process algebra (see e.g. \cite{Pri1}) we consider recursion by constant definitions (as done for the standard $\pi-$calculus e.g. in \cite{MPW}) and not by a bang operator. Moreover, as usual, we assume recursion to be (weakly) guarded. In this way we avoid terms like $A \eqdef A | (\la).P$ for which we would not get a finite cumulative rate for the outgoing Markovian transitions of A.

Let $x,y,\dots$ range over the set $\calx$ of channel names and
$\la$ range over the set $\real^+$ of rate values (representing rates of Markovian distributions).
\begin{definition}\label{MDCalcDef}
The syntax of terms $P, Q, \dots$ is:
$$\begin{array}{lll}
P & ::= & M \mid P | P' \mid (\nu x) P \mid A(x_1, \dots, x_n) \\
M & ::= & \nil \mid \alpha.P  \mid M + M' \\
\alpha & ::= & \overline{x}\lap y \rap \mid x(y) \mid \tau \mid ({\lambda})
\end{array}
$$
We take $\calp$ to be the set of terms $P,Q, \dots$ generated by this syntax.
\end{definition}

We abbreviate $x_1, \dots, x_n$ with $\tilde{x}$ and we assume a given set of constant definitions $A(\tilde{x}) \eqdef P$.
Technically we assume that for every constant $A (\tilde{x}) \eqdef P$ the following holds true. We can, starting from $P$ and by performing a finite number of successive substitutions of constants $B(\tilde{y})$ for their defining term $Q\{\tilde{y}/\tilde{x}\}$ (assuming $B(\tilde{x}) \eqdef Q$) reach a term $P'$ where every constant occurs in $P'$ in the scope of a prefix operator ($A$ is weakly guarded defined, see \cite{MPW}).


\subsection{Labeled Operational Semantics}

Concerning the labeled operational semantics we consider the early semantics of the $\pi-$calculus (see e.g. \cite{SW}) and,
concerning Markovian delays, the classical approach of \cite{Pri1}, i.e. using an explicit (operator dependent) mechanism to distinguish multiple occurrences.

The operational semantics is thus defined by a transition system with two kind of transitions:
\begin{itemize}
\item The classical $\pi-$calculus early semantics labeled transitions, defined in Table~\ref{PiRules} (we assume, as usual, symmetric rules to be also included, the meaning of labels and the definition of functions $n$ (names), $bn$ (bound names) and $fn$ (free names) to be the standard one, see, e.g., \cite{SW} and $\mu,\eta$ to range over sets including all standard labels), and
\item Markovian transitions, defined in Table~\ref{ProofRules},  that
are labeled with pairs $\la,id$ where: $\la$ is the rate of the Markovian delay being executed and $id$ is a unique identifier.
An identifier $id$ is a string over the alphabet $\{ +_0,+_1,|_0,|_1 \}$ and represents the syntactical position of the delay in the syntax tree of the term (the path in the tree). Notice that $\eta$ (differently from $\mu$) also ranges over labels $\la,id$ of Markovian transitions, whose definition, thus, includes also the involved rules of Table~\ref{PiRules}.
\end{itemize}


\begin{table}[t]
$
\begin{array}{ccc}
\multicolumn{3}{c}{
 \tau.P \arrow{\tau} P \hspace{1cm} \overline{x}\lap y \rap .P \arrow{\overline{x} y} P  \hspace{1cm}
x(y).P \arrow{x w} P\{w/y\} } \\[.2cm]
\hspace{1cm} & \infr{ P \arrow{\mu} P'}{P + Q \arrow{\mu} P' } & 
\infr{ P \arrow{\mu} P'}{P | Q \arrow{\mu} P'|Q } \hspace{.3cm}
 bn(\mu) \cap fn(Q) = \emptyset \\
& \infr{ P \arrow{\overline{x}y} P' \hspace{.5cm} Q \arrow{x y} Q'}
{P | Q \arrow{\tau} P'|Q' }  
 & 
\infr{ P \arrow{\overline{x}(y)} P' \hspace{.5cm} Q \arrow{x y} Q'}
{P | Q \arrow{\tau} (\nu x) P'|Q' } \hspace{.3cm} y \notin fn(Q)  \\
\multicolumn{3}{c}{
\infr{ P \arrow{\eta} P'}{(\nu x) P \arrow{\eta} (\nu x) P' } \hspace{.2cm}
 x \notin n(\eta)  
\hspace{.5cm}
\infr{ P \arrow{\overline{x}y} P'}{(\nu y) P \arrow{\overline{x}(y)} P' } \hspace{.2cm}
 y \neq x
\hspace{.5cm}
\infr{ P\{\tilde{y}/\tilde{x}\} \arrow{\eta} P'}{ A(\tilde{y}) \arrow{\eta} P' } \hspace{.2cm}
 A(\tilde{x}) \eqdef P
}
\end{array}
$\\
\caption{Standard Rules}\label{PiRules}
\end{table}

\begin{table}[t]
$
\begin{array}{c}
(\lambda).P \arrow{\la,\varepsilon} P \\[.2cm]
\infr{ P \arrow{\la,id} P'}{P + Q \arrow{\la,+_0 \, id} P' } \hspace{.8cm} 
\infr{ Q \arrow{\la,id} Q'}{P + Q \arrow{\la,+_1 \, id} Q' } \hspace{.8cm}
\infr{ P \arrow{\la,id} P'}{P | Q \arrow{\la,|_0 \, id} P'|Q } \hspace{.8cm} 
\infr{ Q \arrow{\la,id} Q'}{P | Q \arrow{\la,|_1 \, id} P|Q' } \\
\end{array}
$\\
\caption{Identified Markovian Transitions}\label{ProofRules}
\end{table}

The labeled transition system arising from the semantics of a term $P$ has states with both outgoing Markovian
and standard transitions (Interactive Markov Chain) and has to be interpreted as in \cite{Her}.
In particular the Markovian transitions performable in states that also have outgoing $\tau$ transitions are considered to be pre-empted (maximal progress assumption where ``$\tau$'' are instantaneous) and a system is considered to be complete if it cannot undergo further communication with the environment,
i.e. it can perform only $\tau$ and Markovian transitions. 


This is expressed by the notion of Markovian Bisimulation considered in \cite{Her}. We first define the total rate
$\gamma(P,P')$ of transiting from state $P$ to $P'$ with Markovian transitions as
$\gamma(P,P') = \sum_{P \arrow{\la,id} P'} \la$.
This is extended to sets of states in the usual way: $\gamma(P,Set)=\sum_{P' \in Set} \gamma(P,P')$.

We can now consider the Markovian bisimulation as in \cite{Her} (see \cite{Her} for weak bisimulation definition abstracting, when possible, from $\tau$ transitions).

\begin{definition}\label{strongbis}
An equivalence relation $\calb$ on $\calp$ is a strong bisimulation
iff $P \calb Q$ implies for all $\mu$ and all equivalence classes $C$ of $\calb$
\begin{itemize}
\item $P \arrow{\mu} P'$ implies $Q \arrow{\mu} Q'$ for some $Q'$ with $Q' \calb P'$
\item $P \not\!\!\arrow{\tau}$ implies $\gamma(P,C)=\gamma(Q,C)$
\end{itemize}
Two processes $P$ and $Q$ are strongly bisimilar, written $P \sim Q$,if they are
contained in some strong bisimulation.
\end{definition}

\section{Reduction Semantics for Markovian Delays}
\label{MDCalc}

In this section we use our technique based on symbolic representation of rates via stochastic names and stochastic binders
to give a semantics in reduction style for the Markovian $\pi-$calculus of Section \ref{Classical}.

\subsection{Stochastic Names and Stochastic Binders}
\label{MDExt}

We now extend the syntax of terms $P, Q, \dots$ by {\it adding} stochastic names and stochastic binders.
We use 
$q,q',\dots \in \calq\caln$ to range over names used to express symbolically
stochastic information for delays.
$\theta,\theta',\dots$ ranges over $\calx \cup \calq \caln$.

\begin{definition}\label{ExtCalcDef}
We take $\calp_{ext}$ to be the set of terms $P,Q, \dots$ generated by the syntax in Definition \ref{MDCalcDef}, where
we add:
``$(\nu q \sub \la)P$''
to the possible instantiations of $P$, and 
``$(q)$''
to the possible instantiations of $\alpha$ prefixes. 
\end{definition}
$(\nu q \sub \la)$ is a stochastic binder for name $q$ used by a delay, that associates value $\la$ to it.
In the following we use $(\nu x \sub \varepsilon)$ to stand for
$(\nu x)$, i.e. the standard $\pi$-calculus binder, which does not associate any value to $x$, 
This will allow us to write $(\nu \theta \sub \widehat{\,\la})$ to stand for any binder (stochastic or classical) by assuming 
$\widehat{\,\la},\widehat{\,\la'\!}$ to range over $\real^+ \cup \{ \epsilon \}$. 

Notice that we will used $\calp_{ext}$ terms over the extended syntax to define the semantics of $\calp$ terms in 
reduction style. As we will see, for every state of the semantics we will always have a congruent term which belongs to $\calp$.
The definition of structural congruence follows.



\subsection{Structural Congruence Relation}
\label{MDConRel}

\begin{table}[t]
\centerline{$
\begin{array}{llll}
(\nu \theta \sub \pla)P \;| \; Q & \equiv & (\nu \theta \sub \pla)(P | Q)  & \mbox{if} \; \theta \notin fn(Q) \\
(\nu \theta \sub \pla)(\nu \theta' \sub \plap)P & \equiv & 
(\nu \theta' \sub \plap)(\nu \theta \sub \pla)P & \mbox{if} \; \theta \neq \theta' \\
(\nu \theta \sub \pla)\nil & \equiv & 
\nil \\
(M_1 + M_2) + M_3 & \equiv & M_1 + (M_2 + M_3) & \\
M +  N & \equiv & N + M  \\
M + \nil & \equiv & M  \\
(P_1 | P_2) | P_3 & \equiv & P_1 | (P_2 | P_3) & \\
P | Q & \equiv & Q | P  \\
P | \nil & \equiv & P  \\
A(\tilde{y}) & \equiv & P\{\tilde{y}/\tilde{x}\} &  \mbox{if} \; A(\tilde{x}) \eqdef P \\
\end{array}\\[.2cm]
$}
\caption{Standard Structural Congruence Laws (with quantified binders)}\label{ConLaws}
\end{table}

We consider a structural congruence relation over processes in $\calp_{ext}$ defined as usual by a set of laws. We consider the standard laws in Table~\ref{ConLaws} (where we just add the possible presence of quantification in binders)
and, in addition, the following law:
$$
\begin{array}{llll}
(\lambda).P + M& \equiv & (\nu q \sub \la)(\,(q).P + M)  & \mbox{if} \;q \notin fn(M,P) \\
\end{array}
$$
As usual, alpha renaming inside processes, concerning any name $\theta$ and its binder $(\nu \theta \sub \pla)$
is assumed.
Concerning use of parentheses when writing terms in the laws above, we assume binding to take precedence over parallel and prefix to take precedence over sum.


\subsection{Operational Semantics}
\label{MDOpSem}


We now define the semantics of the process algebra of Definition \ref{MDCalc} in reduction style.
As we will see besides standard reduction transitions (defined with standard rules),  we have two other kinds
of transitions representing stochastic execution. Transitions labeled with a stochastic delay name which are treated
similarly as standard reduction transitions and stochastic transitions (that are generated when a stochastic binder is applied at the top level).
The latter lead from a term to a {\it rate distribution} over terms. 

We consider rate distributions $\sigma$ describing how a state can be reached from another one by performing a Markovian transition. 

\begin{definition}
A rate distribution
$\sigma$ over a set $S$ is a {\it non-empty} finite multiset over $\real^+ \times S$, i.e.
a multiset of pairs $(\la, e)$, with $\la \in \real^+$ and $e \in S$. 
We use $RDist(S)$ to denote the set of rate distributions over $S$.
\end{definition}

Due to usage of the structural congruence relation, it will be convenient to use 
rate distributions $\sigma$ over congruence classes of terms, i.e. $RDist(\calp_{ext}/\!\equiv)$.

The operational semantics is defined in terms of three transition relations:
\begin{itemize}
\item Standard reductions~\cite{MilPi} that represent action execution,
leading from a term to another term and denoted by $P \arrow{} P'$.
\item Transitions similar to standard reductions~\cite{MilPi} that represent delay executions and just differ from reductions in that they are additionally labeled with the name $q$ associated to the delay: their quantification is symbolical in that the name of the delay is included and not the actual rate. They are denoted by $P \arrow{q} P'$.
\item Stochastic transitions leading from a term to a rate distribution on congruence classes of terms.
They are denoted denoted by $P \arrow{} \sigma$, where $\sigma \in RDist(\calp_{ext}/\!\equiv)$.
\end{itemize}


\begin{table}[t]
$\begin{array}{c}
(q).P+M \arrow{q} P 
\hspace{1cm}
\mbox{} \tau.P+M \arrow{} P 
\hspace{1cm}
\mbox{} x(z).P+M \; | \; \ov{x}\lap y \rap.Q+N  \arrow{} P\{y/z\} \; | \; Q \\[.4cm]
\infr{ P \arrow{\pq} P'}{P | Q \arrow{\pq} P'|Q } \hspace{.5cm} 
\infr{ P \arrow{\pq} P'}{(\nu \theta \sub \pla) P \arrow{\pq} (\nu \theta \sub \pla) P'} \hspace{.2cm} \theta \neq q
 \hspace{.5cm} 
\infr{ P \equiv Q \hspace{.3cm} Q \arrow{\pq} Q' \hspace{.3cm} Q' \equiv P'}{P \arrow{\pq} P'}
\end{array}$\\
\caption{Standard Rules for Reductions and Symbolically Quantified Transitions}\label{RedRules}
\end{table}

The first two relations are collectively defined as the smallest subset of $\calp_{ext} \times (\calq\caln \cup {\varepsilon}) \times \calp_{ext}$ (where $\varepsilon$ labeled reductions represent the standard ones of~\cite{MilPi}) satisfying the operational rules in Table~\ref{RedRules}, where
$\pq \in \calq\caln \cup \{ \varepsilon \}$.
With respect to the usual ones of~\cite{MilPi}, the rules in Table~\ref{RedRules} only differ for including a rule for delay prefix and for considering names $q \in \calq\caln$ as possible reduction labels (and rates $\lambda$ possibly associated to binders).

%

Stochastic $P \arrow{} \sigma$ transitions are 
defined as the smallest subset of $\calp_{ext} \times RDist(\calp_{ext}/\!\equiv)$ satisfying the operational rules
in Table~\ref{StoRules},
where we assume $(\nu q \sub \la)\sigma = \lmp (\la,[(\nu q \sub \la)P]_{\equiv}) \mid (\la,[P]_{\equiv}) \in \sigma \rmp$.\footnote{We use 
$\lmp$ and $\rmp$ to denote multiset parentheses, $\cup$ to denote multiset union.}
Notice that even if the (Sto2) rule includes a negative
premise the operational semantics is well-defined in that
the inference of transitions can be stratified (see, e.g., \cite{Gro}).

The idea behind the semantics is that $\arrow{} \sigma$ transitions can be generated only for terms
that are (re)organized by the structural congruence in such a way that
stochastic binders (actually binding a delay) are the outermost operators, thus they can correctly evaluate 
the multiset of Markovian transitions leaving a state from symbolic delay transitions (each of which is guaranteed to have a different name).

\begin{table}[t]
$
\begin{array}{cc}
\multicolumn{2}{c}{
 (Sto1) \hspace{.2cm} \infr{P \arrow{} \sigma  \hspace{.5cm} P \arrow{q} P'}
{(\nu q \sub \la) P  \arrow{} (\nu q \sub \la) \sigma \cup \lmp (\la,[(\nu q \sub \la)P']_{\equiv}) \rmp}} \\
 (Sto2) \hspace{.2cm} \infr{P \narrow{} \sigma \hspace{.5cm} P \arrow{q} P'}
{(\nu q \sub \la) P  \arrow{} \lmp (\la,[(\nu q \sub \la)P']_{\equiv}) \rmp}  & \hspace{.5cm}
 (StoCong) \hspace{.2cm} \infr{ P \equiv Q \hspace{.5cm} Q \arrow{} \sigma}{P \arrow{} \sigma}  
\end{array}
$\\
\caption{Rules for Stochastic Transitions}\label{StoRules}
\end{table}


\begin{example}
Consider $P = (5).A | (5).A$ and $A \eqdef (5).A$ we have that $P \equiv (\nu q \sub 5) (q).A | (\nu q \sub 5)(q).A \equiv (\nu q \sub 5)(\nu q' \sub 5) ((q).A | (q').A)$ hence $P$ performs a $\sigma$ transition with $\sigma = \lmp (5,[(5).A | (5).A]_{\equiv}) , (5,[(5).A | (5).A]_{\equiv}) \rmp$, i.e. it is a multiset including two occurrences of the same pair. 
\end{example}

The transition system of terms can be represented in a finitary way by resorting to equivalence
classes $[P]_{\equiv} \in \calp_{ext}/\!\equiv$. This will be important when deriving, as done in Section \ref{Classical} for
the labeled operational semantics, the underlying Interactive Markov Chain.

\begin{definition}\label{TransClass}
Let $P,P' \in \calp_{ext}$. $[P]_{\equiv} \arrow{\pq} [P']_{\equiv}$  whenever $P \arrow{\pq} P'$.
Let $P \in \calp_{ext}$ and $\sigma \in RDist(\calp_{ext}/\!\equiv)$.
$[P]_{\equiv} \arrow{} \sigma$ whenever $P \arrow{} \sigma$.
\end{definition}
The above is well-defined because congruent terms $P$ have the same transitions (rule of closure w.r.t. congruence in each of Tables~\ref{RedRules} and~\ref{StoRules}).
Notice that it is also possible to define directly transitions of equivalence classes, instead of resorting to a definition
like Definition \ref{TransClass}, by putting equivalence classes directly into rules, e.g. for parallel
\cws{8}{4}{
\infr{ [P]_{\equiv} \arrow{\pq} [P']_{\equiv}}{[P | Q]_{\equiv} \arrow{\pq} [P' | Q]_{\equiv} }
}
This would also make the two rules of closure w.r.t. congruence in each of Tables~\ref{RedRules} and~\ref{StoRules} no longer needed (working directly with classes yields the same effect).

Let us now analyze properties of transitions of terms $P \in \calp$ (that is terms belonging to the restricted syntax of Definition~\ref{MDCalcDef}). 

First of all we have that they cannot perform $\arrow{q}$ transitions.

\begin{proposition}
Let $P \in \calp$. There is no $P' \in \calp_{ext}$, $q \in \calq\caln$ such that $P \arrow{q} P'$.
\begin{proof}
A direct consequence of the fact that for any $Q \equiv P$, with $P \in \calp$, we have that, due to the laws of $\equiv$, all delay prefixes $(q)$ occurring in $Q$ are bound (rules for binders do not allow transition labeled with the bound name to be inferred).
\end{proof}
\end{proposition}

The following proposition shows that Markovian transitions are well-defined, i.e. roughly that, no matter the order in which we choose to solve the stochastic binders, the $\sigma$ transition performed is the same.

\begin{proposition}\label{UniqueProp}
Let $P \in \calp$ and $P \arrow{} \sigma$. We have $P \arrow{} \sigma'$ implies $\sigma = \sigma'$.
\begin{proof}
The complete formal proof is quite involved and is presented in Appendix \ref{App}. In the following we just present the structure of the proof in some detail and the main ideas. 

Let $n$ be the number of unguarded syntactical occurrences of
Markovian delay prefixes inside $P$. Formally $n$ is evaluated as follows: $P'$ is obtained from $P$ 
by performing a finite number of successive substitutions of constants so that all constants occur weakly guarded in $P'$
(always possible because recursion is guarded, see Section \ref{MDCalc}); $n$ is the number of Markovian delay prefixes that
occur in $P'$ not in the scope of a prefix operator. 
Since $P \arrow{} \sigma$ we must have $n \geq 1$.

We first show, by induction on $k$, with $0 \leq k \leq n$ ($k=0$ being the base case), that the following statement holds true (see Appendix \ref{App} for details on the induction proof). 
Let $q_1$, \dots, $q_k$ be any set of distinguished names.
There exists $Q$ such that $P \equiv (\nu q_1 \sub \la_1) \dots (\nu q_k \sub \la_k) Q$  (consider $P \equiv Q$ in the case $k =0$) and $\forall i, 1 \leq h \leq k \ldotp Q \arrow{q_h}$.
Moreover for any such $Q$, we have, supposing $\forall h, 1 \leq h \leq k \ldotp Q \arrow{q_h} P'_h$, that:
\begin{itemize}
\item $Q$ includes a unique free unguarded syntactical occurrence of $(q_h)$ $\forall h, 1 \leq h \leq k$ (this reduces to true for the base case $k=0$).
\item If $k<n$, for any $q$ distinguished from $q_h$ $\forall h, 1 \leq h \leq k$, there exist $Q',\la$ such that  $Q \equiv (\nu q \sub \la) Q'$ and $Q' \arrow{q}$, i.e. $Q \arrow{} \sigma$ for some $\sigma$. 
Moreover, any such $Q'$ includes unique free unguarded syntactical occurrences of $(q_h)$ $\forall h, 1 \leq h \leq k$ and,
supposing $\forall h, 1 \leq h \leq k \ldotp Q' \arrow{q_h} P''_h$, we have that $(\nu q \sub \la) P''_h \equiv P'_h$.
\item If, instead, $k=n$ there is no $Q',q,\la$ such that  $Q \equiv (\nu q \sub \la) Q'$ and $Q' \arrow{q}$, i.e. $Q \not \arrow{} \sigma$. 
\end{itemize}

Considering case $k=n$ of the above, we have that, given distinguished names $q_1$, \dots, $q_n$, there exists $P'$ such that $P \equiv (\nu q_1 \sub \la_1) \dots (\nu q_n \sub \la_n) P'$ and $\forall i, 1 \leq h \leq n \ldotp P' \arrow{q_h}$. 
Moreover for any such $P'$ we have that  $P'$ includes a unique free unguarded syntactical occurrence of $(q_i)$ $\forall i, 1 \leq i \leq n$. 

We then show, by induction on the length of inference chains (see Appendix \ref{App}), that inferring a $\sigma$ transition from $P$ (where
$q_1$ is the name of the binder used to infer $\sigma$ in the last rule used in the inference chain and so on and so forth in depth until $q_n$ where the chain of binder applications begins with the (Sto2) rule) entails singling out one of such terms $P'$ 
and that, supposing $\forall h, 1 \leq h \leq n \ldotp P' \arrow{q_h} P'_h$,
the inferred $\sigma$ is such that $$\sigma = \lmp (\la_1,[(\nu q_1 \sub \la_1) \dots (\nu q_n \sub \la_n) P'_1]_{\equiv}) \rmp \cup \dots \cup \lmp (\la_n,[(\nu q_1 \sub \la_1) \dots (\nu q_n \sub \la_n) P'_n]_{\equiv}) \rmp$$
 Notice that this value does 
not depend on the choice of the $P'_h$ such that $P' \arrow{q_h} P'_h$ in that,
since $P'$ includes a unique free unguarded syntactical occurrence of $(q_h)$, possible terms $P'_h$ are all related by $\equiv$  (and, in general,  $P \equiv Q$ implies $(\nu q \sub \la) P \equiv (\nu q \sub \la) Q$ for any $P, Q, q$ and $\la$).

From this fact we can derive that the $\sigma$ inferred is the same for any possible inference chain of a $\sigma$ transition from $P$,
i.e. that no matter the order in which delay prefixes are bound, the same rate
distribution is obtained.
This because, considered $\sigma'$ derived
with another inference chain, such an inference would determine a $P''$ such that $P \equiv (\nu q'_1 \sub \la'_1) \dots (\nu q'_n \sub \la'_n) P''$ and $\forall i, 1 \leq i \leq n \ldotp P'' \arrow{q'_i}$. It is immediate to verify that there exists a permutation $k_i$, with $1 \leq i \leq n$, such that, by alpha-conversion we 
can write $(\nu q'_1 \sub \la'_1) \dots (\nu q'_n \sub \la'_n) P''$ as $(\nu q_{k_1} \sub \la_{k_1}) \dots (\nu q_{k_n} \sub \la_{k_n}) P'''$ where for $P'''$ it holds $P''' \equiv P'$. Thus $\sigma' = \lmp (\la_{k_1},(\nu q_{k_1} \sub \la_{k_1}) \dots (\nu q_{k_n} \sub \la_{k_n}) P''_{k_1}]_{\equiv}) \rmp \cup \dots \cup \lmp (\la_{k_n},[(\nu q_{k_1} \sub \la_{k_1}) \dots (\nu q_{k_n} \sub \la_{k_n}) P''_{k_n}]_{\equiv}) \rmp$, where $\forall i, 1 \leq i \leq n$ we have $P''_{k_i} \equiv P'_{k_i}$
because $P'$ includes a unique free unguarded syntactical occurrence of $(q_i)$ $\forall i, 1 \leq i \leq n$.
$\sigma' = \sigma$ by reordering of the binders, of the members of the multiset union and from the fact that $P \equiv Q$ implies $(\nu q \sub \la) P \equiv (\nu q \sub \la) Q$ for any $P, Q, q$ and $\la$.  

\end{proof}
\end{proposition}

Moreover, from terms in $\calp$ we always reach terms in $\calp$ modulo congruence.

\begin{proposition}
Let $P \in \calp$ and $P' \in \calp_{ext} $ such that $P \arrow{} P'$ or $P \arrow{} \sigma$ and $\exists \lambda$ such that $(\lambda,  [P']_{\equiv}) \in \sigma$. There exists $P'' \in \calp$ such that $P'' \equiv P'$.
\begin{proof}
For any $Q \equiv P$ with $P \in \calp$ we have: all delay prefixes $(q)$ occurring in $Q$ are bound and every stochastic binder can bind at most a single delay prefix $(q)$.
It is easy to verify that this property is preserved by reductions $Q \arrow{} Q'$ or transitions $Q \arrow{} \sigma$ for which $\exists \lambda$ such that $(\lambda,  [Q']_{\equiv}) \in \sigma$, i.e. for any such $Q'$ the same property holds true.
For instance performing the $\sigma$ transition causes the corresponding stochastic binder to not bind any delay prefix
in the target state. 
Any $Q'$ satisfying the property above can be shown to be $\equiv$ to a term of $P$ by just removing stochastic binders not binding any delay prefix and by turning stochastic binder binding a single delay prefix $(q)$ into a Markovian delay
prefix.
\end{proof}
\end{proposition}

A consequence of the above propositions is that, given $P \in \calp$, we can represent the semantics of $P$ as a labeled transition system whose initial state is 
$[P]_{\equiv}$ and, in general, states (congruence classes reachable from $[P]_{\equiv}$) are $[P']_{\equiv}$ such that $P' \in \calp$ (such classes always have a representative in $\calp$). 
In particular, concerning the underlying
Interactive Markov Chain, see Section~\ref{Classical}, we have that 
\begin{itemize}
\item the rate $\gamma([P]_{\equiv},[P']_{\equiv})$ of transiting from state $[P]_{\equiv}$ to state $[P']_{\equiv}$ is: \\ $\sum_{(\la,[P']_{\equiv}) \in \sigma} \la \cdot \mu_{\sigma}(\la,[P']_{\equiv})$ if $[P]_{\equiv} \arrow{} \sigma$ for some $\sigma$; $0$ otherwise\footnote{We assume an empty sum to yield $0$. Given a multiset $m$, $\mu_{m}(e)$ denotes the number of occurrences of element $e$ in $m$.}\\[-.35cm]
\item $\tau$ transitions of the Interactive Markov Chain are standard reductions.
\end{itemize}
The above also provides a definition of bisimulation over states of $P \in \calp$ by applying it to Definition \ref{strongbis} (and to other equivalences in \cite{Her}).
Obviously the reduction semantics yields a closed system interpretation (i.e. corresponding to labeled operational semantics where all actions are assumed to be bound at the top level), thus the obtained Interactive Markov Chain 
is just labeled with $\tau$ actions as for all Interactive Markov Chains corresponding to complete (and not partial) systems, which are commonly the ones to be analyzed for performance.

In the following we formalize this correspondence. In the classical setting this has been done for the $\pi-$calculus, see, e.g.,
Harmony Lemma in \cite{SW} or \cite{MilPi}. Here we extend this result to Markovian delays.



\begin{theorem}[Harmony]
Let $P \in \calp$. We have:
\begin{itemize}
\item $P \arrow{} P'$ iff $P \arrow{\tau} \! \equiv P'$ 
\item $P \arrow{} \sigma$ for some $\sigma$ iff $P \arrow{\la,id} P'$ for some $\la$, $id$ and $P'$. Moreover in the case both hold true, we have that $\forall P' \in \calp$
$$\sum_{(\la,[P']_{\equiv}) \in \sigma} \la \cdot \mu_{\sigma}(\la,[P']_{\equiv}) = \sum_{P \arrow{\la,id} \equiv P'} \la$$
\end{itemize}
\begin{proof} Here we will present in detail the proof of the second item (the first item is standard in that reduction and $\tau$ transitions are derived via standard rules, see, e.g. the Harmony Lemma in \cite{SW}).
Let $n$ be the number of unguarded syntactical occurrences of
Markovian delay prefixes inside $P$ (see proof of Proposition \ref{UniqueProp} 
 for a formal definition of $n$).
We have that 
$P \arrow{} \sigma$ for some $\sigma$  iff $n >0$ (the ``if'' part is obvious, see proof of Proposition \ref{UniqueProp}
for the ``only if'').
Moreover we also obviously have that 
$P \arrow{\la,id} P'$ for some $\la$, $id$ and $P'$ iff $n >0$.

Assumed $n>0$, we consider term $Q$ obtained from $P$ 
by performing a finite number of successive substitutions of constants so that all constants occur weakly guarded in $Q$
(always possible because recursion is guarded, see Section \ref{MDCalc}).
Notice that since $P \equiv Q$ we have $Q \arrow{} \sigma$.
Moreover it is immediate to verify that for each $Q'$ such that
$Q \arrow{\la,id} Q'$ there exists $P'$, with $P' \equiv Q'$, such that $P \arrow{\la,id} P'$;
and viceversa.

Given distinguished names $q_1$, \dots, $q_n$, let $Q'$ and rates $\la_i$, $1 \leq i \leq n$, with $Q'$ including a unique free unguarded syntactical occurrence of each $(q_i)$ $\forall i, 1 \leq i \leq n$, be such that $Q' \{\la_i/q_i \mid 1 \leq i \leq n \} =Q$. We have that $Q \equiv (\nu q_1 \sub \la_1) \dots (\nu q_n \sub \la_n) Q'$ (by extruding stochastic binders up to the top level).
Moreover $\forall h, 1 \leq h \leq n$, let $Q_h$ be $Q' \{ \la_i/q_i \mid 1 \leq i \leq n \wedge i \neq h \}$.
We have that $Q \equiv (\nu q_h \sub \la_h) Q_h$ (by extruding the stochastic binder $q_h$ up to the top level).

We now resort to classical correspondence of transitions between reduction and labeled semantics.
We have: $Q_h \arrow{q_h} Q'_h$ iff  $Q \arrow{\la_h,id_h} Q''_h$ with $Q''_h \! \equiv Q'_h$.
This is proved in a completely analogous way as for the Harmony Lemma in \cite{SW}, since, both unbound $(q_h)$ prefixes for
the reduction semantics in this paper and $(\la)$ for the labeled semantics in this paper have the same behavior as $\tau$ prefixes for standard reduction and labeled semantics (they are just ``consumed'' in the target state with no other effects).

Considered $Q'$ we have that $Q' \arrow{q_h} Q'''_h$ with $Q'_h \equiv (\nu q_1 \sub \la_1) \dots (\nu q_n \sub \la_n) Q'''_h$ (see proof of Proposition \ref{UniqueProp} 
), hence
$\sigma = \lmp (\la_1,[(\nu q_1 \sub \la_1) \dots (\nu q_n \sub \la_n) Q'''_1]_{\equiv}) \rmp \cup \dots \cup \lmp (\la_n,[(\nu q_1 \sub \la_1) \dots (\nu q_n \sub \la_n) Q'''_n]_{\equiv}) \rmp$.
Thus, considered the set $I$ of the indexes $h$, with $1 \leq h \leq n$, such that $[(\nu q_1 \sub \la_1) \dots (\nu q_n \sub \la_n) Q'''_h]_{\equiv} = [P']_{\equiv}$ we have $\sum_{(\la,[P']_{\equiv}) \in \sigma} \la \cdot \mu_{\sigma}(\la,[P']_{\equiv}) = \sum_{h \in I} \la_h$.
Moreover $I$ is also the set of indexes $h$ such that $Q''_h \equiv P'$, because $Q''_h \equiv (\nu q_1 \sub \la_1) \dots (\nu q_n \sub \la_n) Q'''_h$. Therefore, since $h \neq h'$ implies $id_h \neq id_{h'}$,
we also have $\sum_{P \arrow{\la,id} \equiv P'} \la = \sum_{Q \arrow{\la,id} \equiv P'} \la = \sum_{h \in I} \la_h$.
%
\end{proof}
\end{theorem}

\section{Conclusion}

We start by considering (other) related work. As already mentioned this paper is based on a (quite dated) technical report \cite{TechRep} where the idea of stochastic names and stochastic binders has been introduced, but applied to a slightly more complicate Markovian Process Algebra with respect to the one presented here. In particular, while here we consider a process algebra where Markovian delays and standard actions are separated prefixes (by following the simple approach of \cite{Her}), in \cite{TechRep} we apply our technique to a process algebra where Markovian rates are assigned to standard actions,
which become ``durational'' actions (as in \cite{Hil,Pri1,BB}). Such Markovian process algebras are slightly more complex in that the problem of calculating the rate of a synchronized action (out of the rate of the synchronizing actions and the so-called
apparent rates, that is the total rate of the actions of the same type as the synchronizing ones that are collectively performable by one of the processes involved in the synchronization) come into play. The particular way in which such rates are calculated
strictly depends on a fixed structure for the syntax tree for the parallel operators in the term 
(calculating the rates of the synchronizations performable at a certain syntactical level,
for instance in $P\, |\, Q$, from the rates of the transitions performable at lower levels in $P$ and $Q$) and therefore 
is not compatible with a semantics in reduction style. This does not mean that there are no meaningful ways 
for defining rates of synchronized actions in reduction semantics (see \cite{TechRep} and future work below), but different calculations of apparent rates must be adopted that make parallel composition associative \cite{TechRep,KS}, thus yielding different values for the rates. 
The more simplified setting adopted in this paper, instead, allowed us to provide a theorem of consistency with the labeled
operational semantics and to show our approach based on stochastic names and stochastic binders, in its essence, to be correct.
We also would like to mention the approach in the recently appeared paper \cite{CM}. Even if the version of Stochastic $\pi-$calculus presented in \cite{CM} considers a congruence relation on terms, it defines semantics via a labeled operational semantics (and not semantics in reduction style), thus not exploiting congruence while inferring transitions 
(which, for instance, would allow terms to be synchronized to get next to each other and then produce a reduction). In particular \cite{CM} deals with multiplicity following an approach like that of \cite{DLLM}. It is also worth noting that in \cite{CM} 
binders that are very similar to the stochastic binders in this paper (and in \cite{TechRep}) are used, but for a different aim. Here rate values inside binders are used to associate a numerical value to a name representing symbolically a duration for the aim of distinguishing multiple transitions (with the same rate). In \cite{CM}, where multiplicity is dealt with as in \cite{DLLM}, they are used instead to associate rates directly to action names in order to be able to express a novel feature: the ability to pass around actions with a bound name but with an explicit associated rate value.
The similarity in the binder representation would maybe make it possible to apply the technique in this paper to produce 
a reduction semantics for the expressive algebra in \cite{CM} in an overall harmonic approach.

As a final remark, we would like to observe that the approach introduced in this paper, that allows us to independently identify each Markovian delay without introducing an explicit identification mechanism (but by exploiting extrusion and stochastic binders) is applicable also in the case of generally distributed durations where the identification mechanism is necessary to generate clock names (to uniquely relate start of delays with termination of delays), see, for instance, \cite{tcs,BD}. In this case the generally distributed delays would be considered as stochastic binders assigning a general distribution $f$ and immediately binding a delay start prefix followed by a delay termination prefix, both decorated with the same stochastic name that is bound by the binder. 

As future work we consider the application of the technique in this paper to produce  semantics in reduction style in Markovian process algebra with rates associated to actions (using, for instance, the mass action law as in \cite{TechRep} so to have associativity of parallel, see \cite{KS}, or the generative/reactive approach of \cite{BB}, that can be adopted in the context of an associative parallel and whose rate calculation is compatible with inferring transitions via a congruence relation) and in process algebra with probabilistic choices and generally distributed durations \cite{tcs,BD}.


\subparagraph*{Acknowledgements}


We would like to thank Luca Cardelli and Davide Sangiorgi for the fruitful comments and suggestions.










\newpage

\appendix 
\section{Proof of Proposition 8} \label{App}

Let $P \in \calp$ and $P \arrow{} \sigma$. We have $P \arrow{} \sigma'$ implies $\sigma = \sigma'$.
\begin{proof}

Let $n$ be the number of unguarded syntactical occurrences of
Markovian delay prefixes inside $P$. Formally $n$ is evaluated as follows: $P'$ is obtained from $P$ 
by performing a finite number of successive substitutions of constants so that all constants occur weakly guarded in $P'$
(always possible because recursion is guarded, see Section \ref{MDCalc}); $n$ is the number of Markovian delay prefixes that
occur in $P'$ not in the scope of a prefix operator. 
Since $P \arrow{} \sigma$ we must have $n \geq 1$.

We first show, by induction on $k$, with $0 \leq k \leq n$ ($k=0$ being the base case), that the following statement holds true. 
Let $q_1$, \dots, $q_k$ be any set of distinguished names.
There exists $Q$ such that $P \equiv (\nu q_1 \sub \la_1) \dots (\nu q_k \sub \la_k) Q$  (consider $P \equiv Q$ in the case $k =0$) and $\forall i, 1 \leq h \leq k \ldotp Q \arrow{q_h}$.
Moreover for any such $Q$, we have, supposing $\forall h, 1 \leq h \leq k \ldotp Q \arrow{q_h} P'_h$, that:
\begin{itemize}
\item $Q$ includes a unique free unguarded syntactical occurrence of $(q_h)$ $\forall h, 1 \leq h \leq k$ (this reduces to true for the base case $k=0$). This assert (and all other properties of $Q$ stated above) derives directly from the induction hypothesis (case $k-1$): it is
shown by extruding $(\nu q_k \sub \la_k)$ via congruence laws up to the top-level of a term $Q'$ such that $P \equiv (\nu q_1 \sub \la_1) \dots (\nu q_{k-1} \sub \la_{k-1}) Q'$ .
Moreover, since congruence laws preserve uniqueness of free $(q)$ prefixes, we have that $Q$ includes a unique free unguarded syntactical occurrence of $(q_k)$ and unique free unguarded syntactical occurrences of $(q_h)$ $\forall h, 1 \leq h \leq k-1$.
\item If $k<n$, for any $q$ distinguished from $q_h$ $\forall h, 1 \leq h \leq k$, there exist $Q',\la$ such that  $Q \equiv (\nu q \sub \la) Q'$ and $Q' \arrow{q}$, i.e. $Q \arrow{} \sigma$. This is
shown by extruding $(\nu q \sub \la)$ via congruence laws up to the top-level of $Q$.
Moreover, for any such $Q',\la$ we have that $Q'$ includes a unique free unguarded syntactical occurrence of $(q)$ 
and unique free unguarded syntactical occurrences of $(q_h)$ $\forall h, 1 \leq h \leq k$ and, supposing
$\forall h, 1 \leq h \leq k \ldotp Q' \arrow{q_h} P''_h$, we have that $(\nu q \sub \la) P''_h \equiv P'_h$.
This is shown as follows. Since, as already observed, $Q$ includes a unique free unguarded syntactical occurrence of $(q_h)$ $\forall h, 1 \leq h \leq k$ and $Q \equiv (\nu q \sub \la) Q'$, we have the uniqueness results (due to the fact, similarly as in the previous item, that congruence laws preserve uniqueness) and, consequently, that $(\nu q \sub \la) Q' \arrow{q_h} (\nu q \sub \la) P''_h$ with $(\nu q \sub \la) P''_h \equiv P'_h$.
\item If, instead, $k=n$ there is no $Q',q,\la$ such that  $Q \equiv (\nu q \sub \la) Q'$ and $Q' \arrow{q}$, i.e. $Q \not \arrow{} \sigma$. This because, due again to the fact that congruence laws preserve uniqueness, the only occurrences of $(q)$ prefixes in $Q$ are the 
unique free unguarded syntactical occurrences of $(q_h)$ $\forall h, 1 \leq h \leq n$ and in $Q$ there is no
$\la$ such that $Q$ includes an unguarded syntactical occurrence of $(\la)$.
\end{itemize}

Considering case $k=n$ of the above, we have that, given distinguished names $q_1$, \dots, $q_n$, there exists $P'$ such that $P \equiv (\nu q_1 \sub \la_1) \dots (\nu q_n \sub \la_n) P'$ and $\forall i, 1 \leq h \leq n \ldotp P' \arrow{q_h}$. 
Moreover for any such $P'$ we have that  $P'$ includes a unique free unguarded syntactical occurrence of $(q_i)$ $\forall i, 1 \leq i \leq n$. 

In the following we will show, by induction on the length of inference chains, that inferring a $\sigma$ transition from $P$ (where
$q_1$ is the name of the binder used to infer $\sigma$ in the last rule used in the inference chain and so on and so forth in depth until $q_n$ where the chain of binder applications begins with the (Sto2) rule) entails singling out one of such terms $P'$ 
and that, supposing $\forall h, 1 \leq h \leq n \ldotp P' \arrow{q_h} P'_h$,
the inferred $\sigma$ is such that 
$$\sigma = \lmp (\la_1,[(\nu q_1 \sub \la_1) \dots (\nu q_n \sub \la_n) P'_1]_{\equiv}) \rmp \cup \dots \cup \lmp (\la_n,[(\nu q_1 \sub \la_1) \dots (\nu q_n \sub \la_n) P'_n]_{\equiv}) \rmp$$ Notice that this value does 
not depend on the choice of the $P'_h$ such that $P' \arrow{q_h} P'_h$ in that,
since $P'$ includes a unique free unguarded syntactical occurrence of $(q_h)$, possible terms $P'_h$ are all related by $\equiv$  (and, in general,  $P \equiv Q$ implies $(\nu q \sub \la) P \equiv (\nu q \sub \la) Q$ for any $P, Q, q$ and $\la$).

From this fact we can derive that the $\sigma$ inferred is the same for any possible inference chain of a $\sigma$ transition from $P$,
i.e. that no matter the order in which delay prefixes are bound, the same rate
distribution is obtained.
This because, considered $\sigma'$ derived
with another inference chain, such an inference would determine a $P''$ such that $P \equiv (\nu q'_1 \sub \la'_1) \dots (\nu q'_n \sub \la'_n) P''$ and $\forall i, 1 \leq i \leq n \ldotp P'' \arrow{q'_i}$. It is immediate to verify that there exists a permutation $k_i$, with $1 \leq i \leq n$, such that, by alpha-conversion we 
can write $(\nu q'_1 \sub \la'_1) \dots (\nu q'_n \sub \la'_n) P''$ as $(\nu q_{k_1} \sub \la_{k_1}) \dots (\nu q_{k_n} \sub \la_{k_n}) P'''$ where for $P'''$ it holds $P''' \equiv P'$. Thus $\sigma' = \lmp (\la_{k_1},(\nu q_{k_1} \sub \la_{k_1}) \dots (\nu q_{k_n} \sub \la_{k_n}) P''_{k_1}]_{\equiv}) \rmp \cup \dots \cup \lmp (\la_{k_n},[(\nu q_{k_1} \sub \la_{k_1}) \dots (\nu q_{k_n} \sub \la_{k_n}) P''_{k_n}]_{\equiv}) \rmp$, where $\forall i, 1 \leq i \leq n$ we have $P''_{k_i} \equiv P'_{k_i}$
because $P'$ includes a unique free unguarded syntactical occurrence of $(q_i)$ $\forall i, 1 \leq i \leq n$.
$\sigma' = \sigma$ by reordering of the binders, of the members of the multiset union and from the fact that $P \equiv Q$ implies $(\nu q \sub \la) P \equiv (\nu q \sub \la) Q$ for any $P, Q, q$ and $\la$.  

Let us now analyze the structure of the inference chain of a $\sigma$ transition from $P$ by considering an inductive argument
(starting from $k=0$).
Given $k$ such that $0 \leq k \leq n-1$, the structure of any inference sequence of any $\sigma$ transition of $Q$ such that $P \equiv (\nu q_1 \sub \la_1) \dots (\nu q_{k} \sub \la_{k}) Q$ (consider $P \equiv Q$ in the case $k =0$) 
and $\forall h, 1 \leq h \leq k \ldotp Q \arrow{q_h} P'_h$ must be as follows.
Apart from applications of the rule (StoCong), that can be used to get, from $Q \arrow{} \sigma$,
 $(\nu q_{k+1} \sub \la_{k+1}) Q' \arrow{} \sigma$, with $Q'$ such that $Q \equiv (\nu q_{k+1} \sub \la_{k+1}) Q'$ 
and $Q' \arrow{q_{k+1}} P'_{k+1}$ (we have already shown such $Q'$ to always exist for $k \leq n-1$)
the last rule used must be:
\begin{itemize}
\item (Sto2) if $k = n-1$, yielding directly $(\nu q_{n} \sub \la_{n}) Q' \arrow{} \sigma$. This because, as we have shown already, $Q' \not \arrow{} \sigma$. 
The inference sequence is therefore terminated (as far as stochastic binder application is concerned).
\item (Sto1) if $k < n-1$, yielding $(\nu q_{k+1} \sub \la_{k+1}) Q' \arrow \sigma$. This because, as we have shown
already, $Q' \arrow{} \sigma$.
The inference sequence is therefore continued by the inference of a $\sigma'$ transition of term $Q'$. 
\end{itemize}

Hence, by repeatedly applying the statement above ($n$ times starting from $k=0$) we have that, for any $\sigma$ transition of $P$ there exist $Q_1, \dots, Q_n$, with
$P \equiv (\nu q_1 \sub \la_1) \dots (\nu q_{k} \sub \la_{k}) Q_k$ for $1 \leq k \leq n$, such that the inference sequence of the $\sigma$ transition must include $n-1$ usages of the (Sto1)
rule yielding $(\nu q_{1} \sub \la_{1}) Q_1  \arrow{} \sigma_1$, \dots, $(\nu q_{n-1} \sub \la_{n-1}) Q_{n-1}
\arrow{} \sigma_{n-1}$ transitions, with $\sigma_1 = \sigma$,
followed by one usage of the (Sto2) rule yielding $(\nu q_{n} \sub \la_{n}) Q_{n} \arrow{} \sigma_n$ transition.
We have, supposing $\forall h, 1 \leq h \leq n \ldotp Q_n \arrow{q_h} P'_h$,
$\sigma_k = \lmp (\la_{k},[(\nu q_{k} \sub \la_{k}) \dots (\nu q_n \sub \la_n) P'_{k}]_{\equiv}) \rmp \cup \dots \cup \lmp (\la_n,[(\nu q_{k} \sub \la_{k}) \dots (\nu q_n \sub \la_n) P'_n]_{\equiv}) \rmp$.
This is shown by induction on the length of the inference sequence (in terms of number of applications of (Sto1)/(Sto2) rules).
The base case is for $k=n$ (application of the (Sto2) rule). Then we show it holds for $k$ supposing that it holds for $k+1$ (application of the (Sto1) rule).
In the inductive case we exploit the fact, we showed before, that, given $P'''_h$ such that $Q_{k+1} \arrow{q_h} P'''_h$, we have, for $P''_h$ with $Q_k \arrow{q_h} P''_h$, that $(\nu q_{k+1} \sub \la_{k+1}) P'''_h \equiv P''_h$.
\end{proof}

\end{document}